\documentclass[preprint,showpacs,preprintnumbers,amsmath,amssymb]{revtex4}

\usepackage{graphicx}
\usepackage{dcolumn}
\usepackage{bm}
\usepackage{hyperref}
\hypersetup{colorlinks=false}

\begin{document}

\newcommand{\RR}[1]{[#1]}
\newcommand{\intsum}{\sum \kern -15pt \int}
\newfont{\Yfont}{cmti10 scaled 2074}
\newcommand{\Y}{\hbox{{\Yfont y}\phantom.}}
\def\O{{\cal O}}
\newcommand{\bra}[1]{\left< #1 \right| }
\newcommand{\braa}[1]{\left. \left< #1 \right| \right| }
\def\Bra#1#2{{\mbox{\vphantom{$\left< #2 \right|$}}}_{#1}
\kern -2.5pt \left< #2 \right| }
\def\Braa#1#2{{\mbox{\vphantom{$\left< #2 \right|$}}}_{#1}
\kern -2.5pt \left. \left< #2 \right| \right| }
\newcommand{\ket}[1]{\left| #1 \right> }
\newcommand{\kett}[1]{\left| \left| #1 \right> \right.}
\newcommand{\scal}[2]{\left< #1 \left| \mbox{\vphantom{$\left< #1 #2 \right|$}}
\right. #2 \right> }
\def\Scal#1#2#3{{\mbox{\vphantom{$\left<#2#3\right|$}}}_{#1}
{\left< #2 \left| \mbox{\vphantom{$\left<#2#3\right|$}} \right. #3
\right> }}


\title{A Spin-Isospin Dependent 3N Scattering Formalism \\in a 3D Faddeev Scheme}

 \author{M. Harzchi}
\email{mehdi\underline{ }harzchi@khayam.ut.ac.ir}
\author{S. Bayegan}%
 \email{bayegan@khayam.ut.ac.ir}

\affiliation{ Department of Physics, University of Tehran, P.O.Box
14395-547, Tehran, Iran
}%

\date{\today}

\begin{abstract}
We have introduced a spin-isospin dependent three-dimensional
approach for formulation of the three-nucleon scattering. Faddeev
equation is expressed in terms of vector Jacobi momenta and
spin-isospin quantum numbers of each nucleon. Our formalism is based
on connecting the transition amplitude $T$ to momentum-helicity
representations of the two-body $t$-matrix and the deuteron wave
function. Finally the expressions for nucleon-deuteron elastic
scattering and full breakup process amplitudes are presented.

\end{abstract}

\maketitle

\section{Introduction}\label{sec:introduction}

During the past years, the three-dimensional (3D) approach has been
developed for few-body bound and scattering problems [1-15]. The
motivation for developing this approach is introducing a direct
solution of the integral equations avoiding the very involved
angular momentum algebra occurring for the permutations,
transformations and especially for the three-body forces.
Furthermore above the pion production threshold and the GeV region
the three-nucleon (3N) scattering with the 3D formalism is simpler
for more complex few body system by providing a strictly finite
number of coupled 3D integral equation. Conceptually the 3D
formalism considers all partial wave channels automatically.

In the case of the 3N scattering the Faddeev equation has been
formulated for three identical bosons as a function of vector Jacobi
momenta, with the specific stress upon the magnitudes of the momenta
and the angle between them \cite{Elster}, \cite{Schadow}. Adding the
spin-isospin to the 3D formalism is a major additional task, which
will increase more degrees of freedom. This carried out by
Fachruddin \emph{et al.} by considering leading order of Faddeev
equation for breakup process. Recently a 3D formalism based on
operator form has been introduced for formulation of the 3N
scattering \cite{golak}. In this paper we have attempted to
formulate the full breakup and the elastic scattering of
nucleon-deuteron (Nd) by including directly the spin-isospin degrees
of freedom. To this end we have formulated the Faddeev equation for
the 3N scattering with the advantage of using the helicity
representation of NN forces such as the AV18, Bonn-B and the Chiral
potentials~\cite{Fachruddin-PRC62},~\cite{shalchi1}.

This manuscript is organized as follows. In sect.~\ref{sec:Faddeev}
we have derived Faddeev equation in a realistic 3D scheme as a
function of Jacobi momenta vectors and the spin-isospin quantum
numbers. In sect.~\ref{sec:amplitudes} we have derived two
expressions for amplitudes of elastic scattering and breakup process
respectively. Finally in sect.~\ref{sec:summary}, a summary and an
outlook have been presented.

\section{Faddeev equation for the 3N Scattering in the 3D approach} \label{sec:Faddeev}

Faddeev equation for the three identical particle is given
by~\cite{Faddeev}:
\begin{eqnarray}\label{Eq.1}
T|\phi\rangle=tP|\phi\rangle+tG_{0}PT|\phi\rangle,
\end{eqnarray}
where $P=P_{12}P_{23}+P_{13}P_{23}$ is the sum of cyclic and
anti-cyclic permutations of the three nucleons, $t$ denotes the NN
transition matrix determined by a two-body Lippman-Schwinger
equation and $G_{0}$ is the free 3N propagator which is defined as:
\begin{eqnarray}
G_{0}=\frac{1}{E-H_{0}+i\varepsilon},\quad
E=E_{d}+\frac{3}{4m}q_{0}^{2}=E_{d}+\frac{2}{3}E_{lab},\nonumber\\
\end{eqnarray}
where $H_{0}$ and $E$ are the free 3N hamiltonian and the total
energy in the center of mass frame respectively, $E_{d}$ is the
deuteron bounding energy and $\mathbf{q}_{0}$ is the relative
momentum of the projectile nucleon to the deuteron. In order to
solve eq.~(\ref{Eq.1}) in the momentum space we introduce the 3N
free basis states in a 3D formalism
as~\cite{Fachruddin-PRC68},~\cite{Bayegan}:
\begin{eqnarray}
|\mathbf{p}\mathbf{q}\gamma\rangle&\equiv&|\mathbf{p}\mathbf{q}\,m_{s_{1}}
m_{s_{2}}m_{s_{3}}m_{t_{1}}m_{t_{2}}m_{t_{3}}\rangle\equiv|\mathbf{q}
\,m_{s_{1}}m_{t_{1}}\rangle|\mathbf{p}\,m_{s_{2}}m_{s_{3}}m_{t_{2}}m_{t_{3}}\rangle,
\end{eqnarray}
The basis states involve two standard Jacobi momenta $\mathbf{p}$
and $\mathbf{q}$ which are the relative momentum in the subsystem
and the momentum of the spectator with respect to the subsystem
respectively~\cite{Stadler}.
$|\gamma\rangle\equiv|m_{s_{1}}m_{s_{2}}m_{s_{3}}m_{t_{1}}m_{t_{2}}m_{t_{3}}\rangle$
is the spin-isospin parts of the basis states where the quantities
$m_{s_{i}}$ ($m_{t_{i}}$) are the projections of the spin (isospin)
of each three nucleons along its quantization axis. The introduced
basis states are completed and normalized as:
\begin{eqnarray}
&&\sum_{\gamma}\int d\mathbf{p}\int
d\mathbf{q}\,|\mathbf{p}\mathbf{q}\gamma\rangle\langle\mathbf{p}\mathbf{q}\gamma|=
1,\quad\quad\langle\mathbf{p}'\mathbf{q}'\gamma'|\mathbf{p}\mathbf{q}\gamma\rangle=
\delta(\mathbf{p}'-\mathbf{p})\delta(\mathbf{q}'-\mathbf{q})\,\delta_{\gamma'\gamma}.
\end{eqnarray}
Projecting the Faddeev equation on to Jacobi momenta leads to:
\begin{eqnarray}\label{Eq.2}
&&\langle\mathbf{p}\mathbf{q}\gamma|T|\mathbf{q}_{0}\,m^{0}_{s}m^{0}_{t}\Psi^{M_{d}}_{d}
\rangle\nonumber\\&&\hspace{8mm}=\langle\mathbf{p}\mathbf{q}\gamma|tP+tG_{0}PT|\mathbf{q}_{0}\,m^{0}_{s}
m^{0}_{t}\Psi^{M_{d}}_{d}\rangle\nonumber\\&&\hspace{8mm}=\langle\mathbf{p}\mathbf{q}\gamma|tP|
\mathbf{q}_{0}\,m^{0}_{s}m^{0}_{t}\Psi^{M_{d}}_{d}\rangle+\langle\mathbf{p}
\mathbf{q}\gamma|tG_{0}PT|\mathbf{q}_{0}\,m^{0}_{s}m^{0}_{t}\Psi^{M_{d}}_{d}\rangle,
\end{eqnarray}
where:
\begin{eqnarray}
|\phi\rangle\equiv|\mathbf{q}_{0}m^{0}_{s}m^{0}_{t}\Psi^{M_{d}}_{d}\rangle&\equiv&|\mathbf{q}_{0}
m^{0}_{s}m^{0}_{t}\rangle|\Psi^{M_{d}}_{d}\rangle,
\end{eqnarray}
is the initial state where $m^{0}_{s}$ ($m^{0}_{t}$) is spin
(isospin) projection of projectile nucleon along quantization axis
and $|\Psi^{M_{d}}_{d}\rangle$ is the antisymmetrized deuteron state
with $M_{d}$ bing the projection of total angular momentum along
quantization axis. We start by inserting the completeness relation
twice into the first term of eq.~(\ref{Eq.2}):
\begin{eqnarray}\label{Eq.6}
&&\langle\mathbf{p}\mathbf{q}\gamma|\,tP|\mathbf{q}_{0}m^{0}_{s}m^{0}_{t}\Psi^{M_{d}}_{d}
\rangle\nonumber\\&&\hspace{8mm}=\sum_{\gamma'}\int d\mathbf{p}'\int
d\mathbf{q}'\langle\mathbf{p}\mathbf{q}\gamma|t|\mathbf{p}'\mathbf{q}'\gamma'\rangle\nonumber
\sum_{\gamma''}\int d\mathbf{p}''\int
d\mathbf{q}''\langle\mathbf{p}'\mathbf{q}'\gamma'|P|\mathbf{p}''\mathbf{q}''\gamma''\rangle
\nonumber\\&&\hspace{11mm}\times\langle\mathbf{p}''\mathbf{q}''\gamma''|\mathbf{q}_{0}m^{0}_{s}
m^{0}_{t}\Psi^{M_{d}}_{d}\rangle.
\end{eqnarray}
The matrix elements of the permutation operator $P$ are evaluated
as~\cite{Fachruddin-PRC68}:
\begin{eqnarray} \label{Eq.5}
&&\hspace{3mm}\langle\mathbf{p}''\mathbf{q}''\gamma''|P|\mathbf{p}'\mathbf{q}'\gamma'\rangle\nonumber
\\&&=\delta(\mathbf{p}''-\frac{1}{2}\mathbf{q}''-\mathbf{q}')\,\delta(\mathbf{p}'+\mathbf{q}''+\frac{1}{2}
\mathbf{q}')\nonumber\delta_{m''_{s_{1}}m'_{s_{3}}}\delta_{m''_{s_{2}}m'_{s_{1}}}
\delta_{m''_{s_{3}}m'_{s_{2}}}\delta_{m''_{t_{1}}m'_{t_{3}}}\delta_{m''_{t_{2}}m'_{t_{1}}}\delta_{m''_{t_{3}}
m'_{t_{2}}}\nonumber\\&&\hspace{3mm}+\delta(\mathbf{p}''+\frac{1}{2}\mathbf{q}''+\mathbf{q}')
\,\delta(\mathbf{p}'-\mathbf{q}''-\frac{1}{2}\mathbf{q}')\delta_{m''_{s_{1}}
m'_{s_{2}}}\delta_{m''_{s_{2}}m'_{s_{3}}}\delta_{m''_{s_{3}}m'_{s_{1}}}\delta_{m''_{t_{1}}m'_{t_{2}}}
\delta_{m''_{t_{2}}m'_{t_{3}}}\delta_{m''_{t_{3}}m'_{t_{1}}}.
\end{eqnarray}
We have used these relations for the matrix elements of the two-body
$t$-matrix and the deuteron wave function:
\begin{eqnarray}\label{Eq.4}
&&\langle\mathbf{p}\mathbf{q}\gamma|t|\mathbf{p}'\mathbf{q}'\gamma'\rangle=\langle\mathbf{p}m_{s_{2}}m_{s_{3}}m_{t_{2}}m_{t_{3}}|t(z)|\mathbf{p}'m'_{s_{2}}m'_{s_{3}}m'_{t_{2}}
m'_{t_{3}}\rangle\,\delta(\mathbf{q}-\mathbf{q}')\,\delta_{m_{s_{1}}m'_{s_{1}}}
\,\delta_{m_{t_{1}}m'_{t_{1}}},
\end{eqnarray}
\begin{eqnarray}\label{Eq.3}
&&\langle\mathbf{p}''\mathbf{q}''\gamma''|
\mathbf{q}_{0}m^{0}_{s}m^{0}_{t}\Psi^{M_{d}}_{d}\rangle=\langle\mathbf{p}''m''_{s_{2}}m''_{s_{3}}m''_{t_{2}}m''_{t_{3}}|\Psi^{M_{d}}_{d}\rangle\,\delta(\mathbf{q}''-\mathbf{q}_{0})\,\delta_{m''_{s_{1}}m^{0}_{s}}\,\delta_{m''_{t_{1}}
m^{0}_{t}},
\end{eqnarray} where the two-body subsystem energy in a
3N system is $z=E-\frac{3}{4m}q^{2}$. Substituting
eqs.~(\ref{Eq.5}), (\ref{Eq.4}) and (\ref{Eq.3}) into
eq.~(\ref{Eq.6}) yields:
\begin{eqnarray}
&&\hspace{4mm}\langle\mathbf{p}\mathbf{q}\gamma|\,tP|\mathbf{q}_{0}m^{0}_{s}m^{0}_{t}\Psi^{M_{d}}_{d}\rangle
\nonumber\\&&=\sum_{m'_{s_{2}}m'_{t_{2}}}\langle\mathbf{p}m_{s_{2}}m_{s_{3}}m_{t_{2}}m_{t_{3}}|t(z)
|-\mathbf{\pi}m'_{s_{2}}m^{0}_{s}m'_{t_{2}}m^{0}_{t}\rangle\langle\mathbf{\pi}'m_{s_{1}}m'_{s_{2}}m_{t_{1}}m'_{t_{2}}|\Psi^{M_{d}}_{d}\rangle\nonumber \\
&&\hspace{3mm}+\sum_{m'_{s_{3}}m'_{t_{3}}}\langle\mathbf{p}m_{s_{2}}m_{s_{3}}m_{t_{2}}m_{t_{3}}|t(z)|\mathbf{\pi}m^{0}_{s}m'_{s_{3}}m^{0}_{t}m'_{t_{3}}\rangle\langle-\mathbf{\pi}'m'_{s_{3}}m_{s_{1}}m'_{t_{3}}m_{t_{1}}|\Psi^{M_{d}}_{d}\rangle\nonumber
\\&&=\sum_{m'_{s}m'_{t}}\Biggl\{\langle\mathbf{p}m_{s_{2}}m_{s_{3}}m_{t_{2}}m_{t_{3}}|t(z)P_{23}|\mathbf{\pi}m^{0}_{s}m'_{s}m^{0}_{t}m'_{t}\rangle\langle\mathbf{\pi}'m_{s_{1}}m'_{s}m_{t_{1}}m'_{t}|\Psi^{M_{d}}_{d}\rangle\nonumber
\nonumber
\\&&\hspace{14mm}+\langle\mathbf{p}m_{s_{2}}m_{s_{3}}m_{t_{2}}m_{t_{3}}|t(z)|\mathbf{\pi}m^{0}_{s}m'_{s}m^{0}_{t}m'_{t}\rangle\langle\mathbf{\pi}'m_{s_{1}}m'_{s}m_{t_{1}}m'_{t}|P^{-1}_{23}|\Psi^{M_{d}}_{d}\rangle\Biggl\}\nonumber
\\&&=-\sum_{m'_{s}m'_{t}}\langle\mathbf{p}m_{s_{2}}m_{s_{3}}m_{t_{2}}m_{t_{3}}|t(z)(1-P_{23})|\mathbf{\pi}m^{0}_{s}m'_{s}m^{0}_{t}m'_{t}\rangle\langle\mathbf{\pi}'m_{s_{1}}m'_{s}m_{t_{1}}m'_{t}|\Psi^{M_{d}}_{d}\rangle.
\end{eqnarray}
In the last equality we have used the antisymmetry of the deuteron
state $|\Psi^{M_{d}}_{d}\rangle$ and also we have considered:
\begin{eqnarray}
\mathbf{\pi}=\frac{1}{2}\mathbf{q}+\mathbf{q}_{0},\quad\mathbf{\pi}'=\mathbf{q}+\frac{1}{2}\mathbf{q}_{0},
\end{eqnarray}
The antisymmetrized two-body $t$-matrix is given
by~\cite{Fachruddin-PRC62}:
\begin{eqnarray}
\,_{a}\langle\mathbf{p}'m'_{s_{1}}m'_{s_{2}}m'_{t_{1}}m'_{t_{2}}|t|\mathbf{p}m_{s_{1}}m_{s_{2}}m_{t_{1}}m_{t_{2}}\rangle_{a}=\langle\mathbf{p}'m'_{s_{1}}m'_{s_{2}}m'_{t_{1}}m'_{t_{2}}|t(1-P_{12})|\mathbf{p}m_{s_{1}}m_{s_{2}}m_{t_{1}}m_{t_{2}}\rangle,\nonumber\\
\end{eqnarray}
where $|\mathbf{p}m_{s_{1}}m_{s_{2}}m_{t_{1}}m_{t_{2}}\rangle_{a}$
is the antisymmetrized two-body state which is defined as:
\begin{eqnarray}
|\mathbf{p}m_{s_{1}}m_{s_{2}}m_{t_{1}}m_{t_{2}}\rangle_{a}=\frac{1}{\sqrt{2}}(1-P_{12})|\mathbf{p}m_{s_{1}}m_{s_{2}}m_{t_{1}}m_{t_{2}}\rangle.
\end{eqnarray}
Therefore eq.~(11) can be written as:
\begin{eqnarray}
&&\hspace{3mm}\langle\mathbf{p}\mathbf{q}\gamma|tP|\mathbf{q}_{0}m^{0}_{s}m^{0}_{t}\Psi^{M_{d}}_{d}\rangle\nonumber
\\&&=-\sum_{m'_{s}m'_{t}}\,_{a}\langle\mathbf{p}m_{s_{2}}m_{s_{3}}m_{t_{2}}m_{t_{3}}|t(z)|\mathbf{\pi}m^{0}_{s}m'_{s}m^{0}_{t}m'_{t}\rangle_{a}\,\langle
\mathbf{\pi}'m_{s_{1}}m'_{s}m_{t_{1}}m'_{t}|\Psi^{M_{d}}_{d}\rangle.
\end{eqnarray}
Now by inserting the completeness relation twice into the second
term of eq.~(\ref{Eq.2}) and using eqs.~(\ref{Eq.5})
and~(\ref{Eq.4}) we have obtained:
\begin{eqnarray}
&&\hspace{3mm}\langle\mathbf{p}\mathbf{q}\gamma|\,tG_{0}PT|\mathbf{q}_{0}m^{0}_{s}m^{0}_{t}\Psi^{M_{d}}_{d}\rangle\nonumber
\\&&=\sum_{\gamma''}\int
d\mathbf{p}''\int
d\mathbf{q}''\langle\mathbf{p}\mathbf{q}\gamma|t(z)G_{0}|\mathbf{p}''\mathbf{q}''\gamma''\rangle\sum_{\gamma'}\int
d\mathbf{p}'\int
d\mathbf{q}'\langle\mathbf{p}''\mathbf{q}''\gamma''|P|\mathbf{p}'\mathbf{q}'\gamma'\rangle\nonumber
\\&&\hspace{3mm}\times\langle\mathbf{p}'\mathbf{q}'\gamma'|T|\mathbf{q}_{0}m^{0}_{s}m^{0}_{t}\Psi^{M_{d}}_{d}\rangle\nonumber
\\&&=\sum_{m'_{s_{1}}m'_{s_{2}}m'_{t_{1}}m'_{t_{2}}}\int
d\mathbf{q}'\frac{1}{E-\frac{\tilde{\mathbf{\pi}}^{2}}{m}-\frac{3q^{2}}{4m}+i\varepsilon}\langle\mathbf{p}m_{s_{2}}m_{s_{3}}m_{t_{2}}m_{t_{3}}|t(z)|\tilde{\mathbf{\pi}}m'_{s_{1}}m'_{s_{2}}m'_{t_{1}}m'_{t_{2}}\rangle\nonumber
\\&&\hspace{3mm}\times\langle-\tilde{\mathbf{\pi}}'\mathbf{q}'m'_{s_{1}}m'_{s_{2}}m_{s_{1}}m'_{t_{1}}m'_{t_{2}}m_{t_{1}}|T|\mathbf{q}_{0}m^{0}_{s}m^{0}_{t}\Psi^{M_{d}}_{d}\rangle\nonumber
\\&&\hspace{3mm}+\sum_{m'_{s_{1}}m'_{s_{3}}m'_{t_{1}}m'_{t_{3}}}\int d\mathbf{q}'\frac{1}{E-\frac{\tilde{\mathbf{\pi}}^{2}}{m}-\frac{3q^{2}}{4m}+i\varepsilon}\langle\mathbf{p}m_{s_{2}}m_{s_{3}}m_{t_{2}}m_{t_{3}}|t(z)|-\tilde{\mathbf{\pi}}m'_{s_{3}}m'_{s_{1}}m'_{t_{3}}m'_{t_{1}}\rangle\nonumber
\\&&\hspace{3mm}\times\langle\tilde{\mathbf{\pi}}'\mathbf{q}'m'_{s_{1}}m_{s_{1}}m'_{s_{3}}m'_{t_{1}}m_{t_{1}}m'_{t_{3}}|T|\mathbf{q}_{0}m^{0}_{s}m^{0}_{t}\Psi^{M_{d}}_{d}\rangle\nonumber
\nonumber
\\&&=\sum_{m'_{s_{1}}m'_{s}m'_{t_{1}}m'_{t}}\int d\mathbf{q}'\frac{1}{E-\frac{\tilde{\mathbf{\pi}}^{2}}{m}-\frac{3q^{2}}{4m}+i\varepsilon}\Biggl
\{\langle\mathbf{p}m_{s_{2}}m_{s_{3}}m_{t_{2}}m_{t_{3}}|t(z)|\tilde{\mathbf{\pi}}m'_{s_{1}}m'_{s}m'_{t_{1}}m'_{t}\rangle\nonumber
\\&&\hspace{3mm}\times\langle\tilde{\mathbf{\pi}}'\mathbf{q}'m'_{s_{1}}m_{s_{1}}m'_{s}m'_{t_{1}}m_{t_{1}}m'_{t}|P_{23}^{-1}T|\mathbf{q}_{0}m^{0}_{s}m^{0}_{t}\Psi^{M_{d}}_{d}\rangle+\langle\mathbf{p}m_{s_{2}}m_{s_{3}}m_{t_{2}}m_{t_{3}}|t(z)P_{23}|\tilde{\mathbf{\pi}}m'_{s_{1}}m'_{s}m'_{t_{1}}m'_{t}\rangle\nonumber
\\&&\hspace{3mm}\times\langle\tilde{\mathbf{\pi}}'\mathbf{q}'m'_{s_{1}}m_{s_{1}}m'_{s}m'_{t_{1}}m_{t_{1}}m'_{t}|T|\mathbf{q}_{0}m^{0}_{s}m^{0}_{t}\Psi^{M_{d}}_{d}\rangle \Biggl \}\nonumber
\\&&=-\sum_{m'_{s_{1}}m'_{s}m'_{t_{1}}m'_{t}}\int
d\mathbf{q}'\frac{1}{E-\frac{\tilde{\mathbf{\pi}}^{2}}{m}-\frac{3q^{2}}{4m}+i\varepsilon}\times\langle\mathbf{p}m_{s_{2}}m_{s_{3}}m_{t_{2}}m_{t_{3}}|t(z)(1-P_{23})|\tilde{\mathbf{\pi}}m'_{s_{1}}m'_{s}m'_{t_{1}}m'_{t}\rangle\nonumber
\\&&\hspace{3mm}\times\langle\tilde{\mathbf{\pi}}'\mathbf{q}'m'_{s_{1}}m_{s_{1}}m'_{s}m'_{t_{1}}m_{t_{1}}m'_{t}|T|\mathbf{q}_{0}m^{0}_{s}m^{0}_{t}\Psi^{M_{d}}_{d}\rangle\nonumber
\\&&=-\sum_{m'_{s_{1}}m'_{s}m'_{t_{1}}m'_{t}}\int
d\mathbf{q}'\frac{1}{E-\frac{\tilde{\mathbf{\pi}}^{2}}{m}-\frac{3q^{2}}{4m}+i\varepsilon}\,_{a}\langle\mathbf{p}m_{s_{2}}m_{s_{3}}m_{t_{2}}m_{t_{3}}|t(z)|\tilde{\mathbf{\pi}}m'_{s_{1}}m'_{s}m'_{t_{1}}m'_{t}\rangle_{a}\nonumber
\\&&\hspace{3mm}\times\langle\tilde{\mathbf{\pi}}'\mathbf{q}'m'_{s_{1}}m_{s_{1}}m'_{s}m'_{t_{1}}m_{t_{1}}m'_{t}|T|\mathbf{q}_{0}m^{0}_{s}m^{0}_{t}\Psi^{M_{d}}_{d}\rangle,
\end{eqnarray}
where we have considered:
\begin{eqnarray}
\tilde{\mathbf{\pi}}=\frac{1}{2}\mathbf{q}+\mathbf{q}',\quad\tilde{\mathbf{\pi}}'=\mathbf{q}+\frac{1}{2}\mathbf{q}'.
\end{eqnarray}
Final expression for the Faddeev equation is explicitly written as:
\begin{eqnarray}
&&\hspace{3mm}\langle\mathbf{p}\mathbf{q}m_{s_{1}}m_{s_{2}}m_{s_{3}}m_{t_{1}}m_{t_{2}}m_{t_{3}}|T|\mathbf{q}_{0}m^{0}_{s}m^{0}_{t}\Psi^{M_{d}}_{d}\rangle\nonumber
\\&&=-\sum_{m'_{s}m'_{t}}\Biggl\{\,_{a}\langle\mathbf{p}m_{s_{2}}m_{s_{3}}m_{t_{2}}m_{t_{3}}|t(z)|\mathbf{\pi}m^{0}_{s}m'_{s}m^{0}_{t}m'_{t}\rangle_{a}\,\langle
\mathbf{\pi}'m_{s_{1}}m'_{s}m_{t_{1}}m'_{t}|\Psi^{M_{d}}_{d}\rangle\nonumber
\\&&\hspace{3mm}+\sum_{m'_{s_{1}}m'_{t_{1}}}\int
d\mathbf{q}'\,_{a}\langle\mathbf{p}m_{s_{2}}m_{s_{3}}m_{t_{2}}m_{t_{3}}|t(z)|\tilde{\mathbf{\pi}}m'_{s_{1}}m'_{s}m'_{t1_{}}m'_{t}\rangle_{a}\nonumber
\\&&\hspace{3mm}\times\frac{\langle\tilde{\mathbf{\pi}}'\mathbf{q}'m'_{s_{1}}m_{s_{1}}m'_{s}m'_{t_{1}}m_{t_{1}}m'_{t}|T|\mathbf{q}_{0}m^{0}_{s}m^{0}_{t}\Psi^{M_{d}}_{d}\rangle}{E-\frac{1}{m}(q^{2}+\mathbf{q}\cdot
\mathbf{q}'+q'^{2})+i\varepsilon}\Biggl\}.
\end{eqnarray}
Since the transition operator $T$ is needed for all of $\mathbf{q}$
values and We know that the two-body interaction supports a bound
state, which is characterized by a pole in the two-body $t$-matrix
at the two-body binding energy $E_{d}$, thus we need to consider
this pole, which is located at $z=E_{d}$. The residue at the pole
can be explicitly extracted by defining:
\begin{eqnarray}
\hat{t}^{a}=(z-E_{d})t^{a},
\end{eqnarray}
Since the pole of $t^{a}$ will be present in $T$, we define:
\begin{eqnarray}
\hat{T}=(z-E_{d})T,
\end{eqnarray}
Therefore eq.~(18) can be written as:
\begin{eqnarray}
&&\hspace{3mm}\langle\mathbf{p}\mathbf{q}m_{s_{1}}m_{s_{2}}m_{s_{3}}m_{t_{1}}m_{t_{2}}m_{t_{3}}|\hat{T}|\mathbf{q}_{0}m^{0}_{s}m^{0}_{t}\Psi^{M_{d}}_{d}\rangle\nonumber \\
&&=-\sum_{m'_{s}m'_{t}}\Biggl\{\,_{a}\langle\mathbf{p}m_{s_{2}}m_{s_{3}}m_{t_{2}}m_{t_{3}}|\,\hat{t}(z)|\mathbf{\pi}m^{0}_{s}m'_{s}m^{0}_{t}m'_{t}\rangle_{a}\,\langle
\mathbf{\pi}'m_{s_{1}}m'_{s}m_{t_{1}}m'_{t}|\Psi^{M_{d}}_{d}\rangle\nonumber\\
&&\hspace{3mm}+\sum_{m'_{s_{1}}m'_{t_{1}}}\int
d\mathbf{q}'\frac{\,_{a}\langle\mathbf{p}m_{s_{2}}m_{s_{3}}m_{t_{2}}m_{t_{3}}|\,\hat{t}(z)|\tilde{\mathbf{\pi}}m'_{s_{1}}m'_{s}m'_{t1_{}}m'_{t}\rangle_{a}}{E-\frac{1}{m}(q^{2}+\mathbf{q}\cdot
\mathbf{q}'+q'^{2})+i\varepsilon}\nonumber \\
&&\hspace{3mm}\times\frac{\langle\tilde{\mathbf{\pi}}'\mathbf{q}'m'_{s_{1}}m_{s_{1}}m'_{s}m'_{t_{1}}m_{t_{1}}m'_{t}|\hat{T}|\mathbf{q}_{0}m^{0}_{s}m^{0}_{t}\Psi^{M_{d}}_{d}\rangle}{E-\frac{3}{4m}q'^{2}-E_{d}+i\varepsilon}\Biggl\}.
\end{eqnarray}
As a simplification we rewrite eq.~(21) as:
\begin{eqnarray}
&&\hspace{3mm}\hat{T}^{m^{0}_{s}m^{0}_{t},M_{d}}_{m_{s_{1}}m_{s_{2}}m_{s_{3}}m_{t_{1}}m_{t_{2}}m_{t_{3}}}(\mathbf{p},\mathbf{q};\mathbf{q}_{0})\nonumber\\
&&=-\sum_{m'_{s}m'_{t}}\Biggl\{\hat{t}_{a}\,_{m_{s_{2}}m_{s_{3}}m_{t_{2}}m_{t_{3}}}^{m^{0}_{s}m'_{s}m^{0}_{t}m'_{t}}(\mathbf{p},\mathbf{\pi};z)\,\,\Psi^{M_{d}}_{m_{s_{1}}m'_{s}m_{t_{1}}m'_{t}}(\mathbf{\pi}')\nonumber\\
&&\hspace{3mm}+\sum_{m''_{s}m''_{t}}\int
d\mathbf{q}''\frac{\hat{t}_{a}\,_{m_{s_{2}}m_{s_{3}}m_{t_{2}}m_{t_{3}}}^{m''_{s}m'_{s}m''_{t}m'_{t}}(\mathbf{p},\tilde{\mathbf{\pi}},z)}{E-\frac{1}{m}(q^{2}+\mathbf{q}\cdot
\mathbf{q}''+q''^{2})+i\varepsilon}\frac{\hat{T}^{m^{0}_{s}m^{0}_{t},M_{d}}_{m''_{s}m_{s_{1}}m'_{s}m''_{t}m_{t_{1}}m'_{t}}(\tilde{\mathbf{\pi}}',\mathbf{q}'';\mathbf{q}_{0})}{E-\frac{3}{4m}q''^{2}-E_{d}+i\varepsilon}\Biggl\}.
\end{eqnarray}
For solving this integral equation one needs the matrix elements of
the deuteron wave function and the antisymmetrized two-body
$t$-matrix. We have connected these to their helicity
representations in appendices A and B respectively.

\section{Derivation of the 3D expressions for the 3N elastic and breakup scattering amplitudes}\label{sec:amplitudes}

\subsection{Elastic scattering}
In the Faddeev scheme the operator $U$ for the Nd elastic scattering
is defined by:
\begin{eqnarray}
U=PG^{-1}_{0}+PT,
\end{eqnarray}
Differential cross section for Nd elastic scattering in the center
of mass frame is given by:
\begin{eqnarray}
\frac{d\sigma}{d\Omega}=(2\pi)^{4}(\frac{2m}{3})^{2}\frac{1}{6}\sum_{m^{0}_{s}m_{s}M_{d}M'_{d}}|U^{M'_{d},M_{d}}_{m_{s}m_{t},m^{0}_{s}m^{0}_{t}}(\mathbf{q},\mathbf{q}_{0})|^{2},
\end{eqnarray}
where
$U^{M'_{d},M_{d}}_{m_{s}m_{t},m^{0}_{s}m^{0}_{t}}(\mathbf{q},\mathbf{q}_{0})$
is the elastic scattering amplitude which is defined as:
\begin{eqnarray} \label{Eq.7}
&&\hspace{3mm}U^{M'_{d},M_{d}}_{m_{s}m_{t},m^{0}_{s}m^{0}_{t}}(\mathbf{q},\mathbf{q}_{0})\nonumber\\&&\equiv\langle\mathbf{q}m_{s}m_{t}\Psi^{M'_{d}}_{d}|U|\mathbf{q}_{0}m^{0}_{s}m^{0}_{t}\Psi^{M_{d}}_{d}\rangle\nonumber \\
&&=\langle\mathbf{q}m_{s}m_{t}\Psi^{M'_{d}}_{d}|(PG^{-1}_{0}+PT)|\mathbf{q}_{0}m^{0}_{s}m^{0}_{t}\Psi^{M_{d}}_{d}\rangle\nonumber \\
&&=\langle\mathbf{q}m_{s}m_{t}\Psi^{M'_{d}}_{d}|PG^{-1}_{0}|\mathbf{q}_{0}m^{0}_{s}m^{0}_{t}\Psi^{M_{d}}_{d}\rangle+\langle\mathbf{q}m_{s}m_{t}\Psi^{M'_{d}}_{d}|PT|\mathbf{q}_{0}m^{0}_{s}m^{0}_{t}\Psi^{M_{d}}_{d}\rangle,
\end{eqnarray}
where:
\begin{eqnarray}
|\mathbf{q}_{0}m^{0}_{s}m^{0}_{t}\Psi^{M_{d}}_{d}\rangle\equiv|\mathbf{q}_{0}m^{0}_{s}m^{0}_{t}\rangle|\Psi^{M_{d}}_{d}\rangle,\quad\quad\mathbf{q}m_{s}m_{t}\Psi^{M'_{d}}_{d}|\equiv\langle\mathbf{q}m_{s}m_{t}|\langle\Psi^{M'_{d}}_{d}|,
\end{eqnarray}
are the initial and the final states respectively. For derivation of
an expression for the elastic scattering amplitude we start by
inserting the completeness relation twice into the first term of
eq.~(\ref{Eq.7}) as:
\begin{eqnarray} \label{Eq.8}
&&\hspace{3mm}\langle\mathbf{q}m_{s}m_{t}\Psi^{M'_{d}}_{d}|PG^{-1}_{0}|\mathbf{q}_{0}m^{0}_{s}m^{0}_{t}\Psi^{M_{d}}_{d}\rangle\nonumber\\&&=\sum_{\gamma''}\int
d\mathbf{p}''\int
d\mathbf{q}''\langle\mathbf{q}m_{s}m_{t}\Psi^{M'_{d}}_{d}|\mathbf{p}''\mathbf{q}''\gamma''\rangle\sum_{\gamma'}\int
d\mathbf{p}'\int
d\mathbf{q}'(E-\frac{p'^{2}}{m}-\frac{3q'^{2}}{4m})\nonumber\\&&\hspace{3mm}\times\langle\mathbf{p}''\mathbf{q}''\gamma''|P|\mathbf{p}'\mathbf{q}'\gamma'\rangle\langle\mathbf{p}'\mathbf{q}'\gamma'|\mathbf{q}_{0}m^{0}_{s}m^{0}_{t}\Psi^{M_{d}}_{d}\rangle.
\end{eqnarray}
with considering eq.~(\ref{Eq.5}) and following relations:
\begin{eqnarray}\label{Eq.9}
&&\hspace{3mm}\langle\mathbf{q}m_{s}m_{t}\Psi^{M'_{d}}_{d}|\mathbf{p}''\mathbf{q}''\gamma''\rangle=\langle\Psi^{M'_{d}}_{d}|\mathbf{p}''m''_{s_{2}}m''_{s_{3}}m''_{t_{2}}m''_{t_{3}}\rangle\delta(\mathbf{q}-\mathbf{q}'')\,\delta_{m_{s}m''_{s_{1}}}\delta_{m_{t}m''_{t_{1}}},
\end{eqnarray}
\begin{eqnarray}\label{Eq.10}
&&\hspace{3mm}\langle\mathbf{p}'\mathbf{q}'\gamma'|\mathbf{q}_{0}m^{0}_{s}m^{0}_{t}\Psi^{M_{d}}_{d}\rangle=\langle\mathbf{p}'m'_{s_{2}}m'_{s_{3}}m'_{t_{2}}m'_{t_{3}}|\Psi^{M_{d}}_{d}\rangle\delta(\mathbf{q}'-\mathbf{q}_{0})\,\delta_{m'_{s_{1}}m^{0}_{s}}\delta_{m'_{t_{1}}m^{0}_{t}},
\end{eqnarray}
Equation~(\ref{Eq.8}) can be written as:
\begin{eqnarray}
&&\hspace{3mm}\langle\mathbf{q}m_{s}m_{t}\Psi^{M'_{d}}_{d}|PG^{-1}_{0}|\mathbf{q}_{0}m^{0}_{s}m^{0}_{t}\Psi^{M_{d}}_{d}\rangle\nonumber\\
&&=(E-\frac{\pi'^{2}}{m}-\frac{3q_{0}^{2}}{4m})\sum_{m'_{s_{2}}m'_{t_{2}}}\langle\Psi^{M'_{d}}_{d}|
\pi
m^{0}_{s}m'_{s_{2}}m^{0}_{t}m'_{t_{2}}\rangle\langle-\pi'm'_{s_{2}}m_{s}m'_{t_{2}}m_{t}|\Psi^{M_{d}}_{d}\rangle\nonumber\\&&\hspace{3mm}+\sum_{m'_{s_{3}}m'_{t_{3}}}\langle\Psi^{M'_{d}}_{d}|
-\pi
m'_{s_{3}}m^{0}_{s}m'_{t_{3}}m^{0}_{t}\rangle\langle\pi'm_{s}m'_{s_{3}}m_{t}m'_{t_{3}}|\Psi^{M_{d}}_{d}\rangle
\nonumber\\&&=(E-\frac{\pi'^{2}}{m}-\frac{3q_{0}^{2}}{4m})\sum_{m'_{s}m'_{t}}\Biggl\{\langle\Psi^{M'_{d}}_{d}|
\pi
m^{0}_{s}m'_{s}m^{0}_{t}m'_{t}\rangle\langle\pi'm_{s}m'_{s}m_{t}m'_{t}|P^{-1}_{23}|\Psi^{M_{d}}_{d}\rangle\nonumber\\&&\hspace{3mm}+\langle\Psi^{M'_{d}}_{d}|P_{23}|
\pi
m^{0}_{s}m'_{s}m^{0}_{t}m'_{t}\rangle\langle\pi'm_{s}m'_{s}m_{t}m'_{t}|\Psi^{M_{d}}_{d}\rangle\Biggl\}\nonumber
\\&&=-2(E-\frac{\pi'^{2}}{m}-\frac{3q_{0}^{2}}{4m})\sum_{m'_{s}m'_{t}}\langle\Psi^{M'_{d}}_{d}|
\pi
m^{0}_{s}m'_{s}m^{0}_{t}m'_{t}\rangle\langle\pi'm_{s}m'_{s}m_{t}m'_{t}|\Psi^{M_{d}}_{d}\rangle\nonumber
\\&&=-2(E-\frac{1}{m}(q^{2}+\mathbf{q}\cdot\mathbf{q}_{0}+q^{2}_{0}))\sum_{m'_{s}m'_{t}}\langle\Psi^{M'_{d}}_{d}|
\pi
m^{0}_{s}m'_{s}m^{0}_{t}m'_{t}\rangle\langle\pi'm_{s}m'_{s}m_{t}m'_{t}|\Psi^{M_{d}}_{d}\rangle.
\end{eqnarray}
Now by inserting the completeness relation twice into the second
term of eq.~(\ref{Eq.7}) and using eqs.~(\ref{Eq.5}) and
(\ref{Eq.9})
 we have obtained:
\begin{eqnarray} \label{eq.Vlowk}
&&\hspace{3mm}\langle\mathbf{q}m_{s}m_{t}\Psi^{M'_{d}}_{d}|PT|\mathbf{q}_{0}m^{0}_{s}m^{0}_{t}\Psi^{M_{d}}_{d}\rangle\nonumber\\
&&=\sum_{\gamma''}\int d\mathbf{p}''\int
d\mathbf{q}''\langle\mathbf{q}m_{s}m_{t}\Psi^{M'_{d}}_{d}|\mathbf{p}''\mathbf{q}''\gamma''\rangle
\sum_{\gamma'}\int d\mathbf{p}'\int
d\mathbf{q}'\langle\mathbf{p}''\mathbf{q}''\gamma''|P|\mathbf{p}'\mathbf{q}'\gamma'\rangle\nonumber\\&&\hspace{3mm}\times\langle\mathbf{p}'\mathbf{q}'\gamma'|T|\mathbf{q}_{0}m^{0}_{s}m^{0}_{t}\Psi^{M_{d}}_{d}\rangle\nonumber\\&&=\sum_{m'_{s_{1}}m'_{t_{1}}}\int
d\mathbf{q}'
\Biggl\{\sum_{m'_{s_{2}}m'_{t_{2}}}\langle\Psi^{M'_{d}}_{d}|
\tilde{\pi}m'_{s_{1}}m'_{s_{2}}m'_{t_{1}}m'_{t_{2}}\rangle\langle-\tilde{\mathbf{\pi}}'\mathbf{q}'m'_{s_{1}}m'_{s_{2}}m_{s}m'_{t_{1}}m'_{t_{2}}m_{t}|T|\mathbf{q}_{0}m^{0}_{{s}}m^{0}_{{t}}\Psi^{M_{d}}_{d}\rangle\nonumber\\
&&\hspace{3mm}+\sum_{m'_{s_{3}}m'_{t_{3}}}\langle\Psi^{M'_{d}}_{d}|
-\tilde{\mathbf{\pi}}m'_{s_{3}}m'_{s_{1}}m'_{t_{3}}m'_{t_{1}}\rangle\langle\tilde{\mathbf{\pi}}'\mathbf{q}'m'_{s_{1}}m_{s}m'_{s_{3}}m'_{t_{1}}m_{t}m'_{t_{3}}|T|\mathbf{q}_{0}m^{0}_{{s}}m^{0}_{{t}}\Psi^{M_{d}}_{d}\rangle\Biggl\}\nonumber\\
&&=\sum_{m'_{s_{1}}m'_{t_{1}}m'_{s}m'_{t}}\int
d\mathbf{q}'\Biggl\{\langle\Psi^{M'_{d}}_{d}|
\tilde{\mathbf{\pi}}m'_{s_{1}}m'_{s}m'_{t_{1}}m'_{t}\rangle\langle\tilde{\mathbf{\pi}}'\mathbf{q}'m'_{s_{1}}m_{s}m'_{s}m'_{t_{1}}m_{t}m'_{t}|P_{23}^{-1}T|\mathbf{q}_{0}m^{0}_{{s}}m^{0}_{{t}}\Psi^{M_{d}}_{d}\rangle\nonumber\\
&&\hspace{3mm}+\langle\Psi^{M'_{d}}_{d}|P_{23}|
\tilde{\mathbf{\pi}}m'_{s_{1}}m'_{s}m'_{t_{1}}m'_{t}\rangle\langle\tilde{\mathbf{\pi}}'\mathbf{q}'m'_{s_{1}}m_{s}m'_{s}m'_{t_{1}}m_{t}m'_{t}|T|\mathbf{q}_{0}m^{0}_{{s}}m^{0}_{{t}}\Psi^{M_{d}}_{d}\rangle\Biggl\}\nonumber\\
&&=-2\sum_{m'_{s}m'_{t}m'_{s_{1}}m'_{t_{1}}}\int
d\mathbf{q}'\langle\Psi^{M'_{d}}_{d}|
\tilde{\mathbf{\pi}}m'_{s_{1}}m'_{s}m'_{t_{1}}m'_{t}\rangle\langle\tilde{\mathbf{\pi}}'\mathbf{q}'m'_{s_{1}}m_{s}m'_{s}m'_{t_{1}}m_{t}m'_{t}|T|\mathbf{q}_{0}m^{0}_{{s}}m^{0}_{{t}}\Psi^{M_{d}}_{d}\rangle
\nonumber \\&&= -2\sum_{m'_{s}m'_{t}m'_{s_{1}}m'_{t_{1}}}\int
d\mathbf{q}'\langle\Psi^{M'_{d}}_{d}|
\tilde{\pi}m'_{s_{1}}m'_{s}m'_{t_{1}}m'_{t}\rangle\frac{\langle\tilde{\mathbf{\pi}}'\mathbf{q}'m'_{s_{1}}m_{s}m'_{s}m'_{t_{1}}m_{t}m'_{t}|\hat{T}|\mathbf{q}_{0}m^{0}_{{s}}m^{0}_{{t}}\Psi^{M_{d}}_{d}\rangle}{E-\frac{3}{4m}q'^{2}-E_{d}+i\varepsilon}.
\end{eqnarray}
Final expression for the matrix elements of elastic scattering
amplitude has been obtained as:
\begin{eqnarray} \label{Eq.11}
&&\hspace{3mm}\langle\mathbf{q}m_{s}m_{t}\Psi^{M'_{d}}_{d}|U|\mathbf{q}_{0}m^{0}_{s}m^{0}_{t}\Psi^{M_{d}}_{d}\rangle\nonumber\\
&&=
-2\sum_{m'_{s}m'_{t}}\Biggl\{(E-\frac{1}{m}(q^{2}+\mathbf{q}\cdot\mathbf{q}_{0}+q^{2}_{0}))\langle\Psi^{M'_{d}}_{d}|
\mathbf{\pi}m^{0}_{s}m'_{s}m^{0}_{t}m'_{t}\rangle\langle\mathbf{\pi}'m_{s}m'_{s}m_{t}m'_{t}|\Psi^{M_{d}}_{d}\rangle\nonumber
\nonumber\\&&\hspace{3mm}+\sum_{m'_{s_{1}}m'_{t_{1}}}\int
d\mathbf{q}'\langle\Psi^{M'_{d}}_{d}|
\tilde{\mathbf{\pi}}m'_{s_{1}}m'_{s}m'_{t_{1}}m'_{t}\rangle\frac{\langle\tilde{\mathbf{\pi}}'\mathbf{q}'m'_{s_{1}}m_{s}m'_{s}m'_{t_{1}}m_{t}m'_{t}|\hat{T}|\mathbf{q}_{0}m^{0}_{{s}}m^{0}_{{t}}\Psi^{M_{d}}_{d}\rangle}{E-\frac{3}{4m}q'^{2}-E_{d}+i\varepsilon}\Biggl\}.
\end{eqnarray}
As a simplification we rewrite eq.~(\ref{Eq.11}) as:
\begin{eqnarray} \label{eq.Vlowk}
&&\hspace{3mm}U_{m_{s}m_{t},m^{0}_{s}m^{0}_{t}}^{M'_{d},M_{d}}(\mathbf{q},\mathbf{q}_{0})
\nonumber\\
&&=-2\sum_{m'_{s}m'_{t}}\Biggl\{(E-\frac{1}{m}(q^{2}+\mathbf{q}\cdot\mathbf{q}_{0}+q^{2}_{0}))\,\,\Psi^{\ast
M'_{d}}_{m^{0}_{s}m'_{s}m^{0}_{t}m'_{t}}(\mathbf{\pi})\nonumber\\&&\hspace{3mm}\times\Psi^{M_{d}}_{m_{s}m'_{s}m_{t}m'_{t}}(\mathbf{\pi}')
+\sum_{m''_{s}m''_{t}}\int d\mathbf{q}''\,\,\Psi^{\ast
M'_{d}}_{m''_{s}m'_{s}m''_{t}m'_{t}}
(\tilde{\mathbf{\pi}})\frac{\hat{T}^{m^{0}_{{s}}m^{0}_{{t}},M_{d}}_{m''_{s}m_{s}m'_{s}m''_{t}m_{t}m'_{t}}(\tilde{\mathbf{\pi}}',\mathbf{q}'';\mathbf{q}_{0})}{E-\frac{3}{4m}q''^{2}-E_{d}+i\varepsilon}\Biggl\}.
\end{eqnarray}
\subsection{Full breakup process}
 In this stage we have derived an expression for the matrix
elements of the breakup process. In the Faddeev scheme the operator
$U_{0}$ for the Nd breakup process is given by:
\begin{eqnarray}
U_{0}=(1+P)T.
\end{eqnarray}
Differential cross section for the Nd breakup process is defined
by~\cite{Fachruddin-PRC68}:
\begin{eqnarray}
\frac{d\sigma}{d\Omega_{q}dq}=(2\pi)^{4}\frac{m^{2}}{3q_{0}}p\,q^{2}\frac{1}{6}\sum_{M_{d}m^{0}_{s}\gamma}\int
d\mathbf{\hat{p}}|U^{M_{d}m^{0}_{s},\gamma}_{0}(\mathbf{p},\mathbf{q};\mathbf{q}_{0})|^{2},
\end{eqnarray}
where the matrix elements of the full breakup amplitude
$U^{M_{d}m^{0}_{s},\gamma}_{0}(\mathbf{p},\mathbf{q};\mathbf{q}_{0})$
is defined as:
\begin{eqnarray}\label{Eq.12}
U^{M_{d}m^{0}_{s},\gamma}_{0}(\mathbf{p},\mathbf{q};\mathbf{q}_{0})&\equiv&\langle\mathbf{p}\mathbf{q}\gamma|U_{0}|\mathbf{q}_{0}m^{0}_{s}m^{0}_{t}\Psi^{M_{d}}_{d}\rangle\nonumber \\
&=&\langle\mathbf{p}\mathbf{q}\gamma|(1+P)T|\mathbf{q}_{0}m^{0}_{s}m^{0}_{t}\Psi^{M_{d}}_{d}\rangle
+\langle\mathbf{p}\mathbf{q}\gamma|T|\mathbf{q}_{0}m^{0}_{s}m^{0}_{t}\Psi^{M_{d}}_{d}\rangle
\nonumber\\&&+\langle\mathbf{p}\mathbf{q}\gamma|(P_{12}P_{23})T|\mathbf{q}_{0}m^{0}_{s}m^{0}_{t}\Psi^{M_{d}}_{d}\rangle+\langle\mathbf{p}\mathbf{q}\gamma|(P_{13}P_{23})T|\mathbf{q}_{0}m^{0}_{s}m^{0}_{t}\Psi^{M_{d}}_{d}\rangle,\nonumber\\
\end{eqnarray}
where $|\mathbf{q}_{0}m^{0}_{s}m^{0}_{t}\Psi^{M_{d}}_{d}\rangle$ and
$\langle\mathbf{p}\mathbf{q}\gamma|$ are initial and final states
respectively. By applying the permutation operator $P_{12}P_{23}$
and $P_{13}P_{23}$ to the final state, eq.~(\ref{Eq.12}) can be
written as~\cite{Fachruddin-PRC68}:
\begin{eqnarray}\label{Eq.13}
&&\hspace{4mm}U^{M_{d}m^{0}_{s},\gamma}_{0}(\mathbf{p},\mathbf{q};\mathbf{q}_{0})\nonumber\\&&\equiv\langle\mathbf{p}\mathbf{q}m_{s_{1}}m_{s_{2}}m_{s_{3}}m_{t_{1}}m_{t_{2}}m_{t_{3}}|T|\mathbf{q}_{0}m^{0}_{s}m^{0}_{t}\Psi^{M_{d}}_{d}\rangle\nonumber \\
&&+\langle(-\frac{1}{2}\mathbf{p}-\frac{3}{4}\mathbf{q})(\mathbf{p}-\frac{1}{2}\mathbf{q})m_{s_{2}}m_{s_{3}}m_{s_{1}}m_{t_{2}}m_{t_{3}}m_{t_{1}}|T|\mathbf{q}_{0}m^{0}_{s}m^{0}_{t}\Psi^{M_{d}}_{d}\rangle\nonumber \\
&&+\langle(-\frac{1}{2}\mathbf{p}+\frac{3}{4}\mathbf{q})(-\mathbf{p}-\frac{1}{2}\mathbf{q})m_{s_{3}}m_{s_{1}}m_{s_{2}}m_{t_{3}}m_{t_{1}}m_{t_{2}}|T|\mathbf{q}_{0}m^{0}_{s}m^{0}_{t}\Psi^{M_{d}}_{d}\rangle\nonumber
\\&&=\frac{\langle\mathbf{p}\mathbf{q}m_{s_{1}}m_{s_{2}}m_{s_{3}}m_{t_{1}}m_{t_{2}}m_{t_{3}}|\hat{T}|\mathbf{q}_{0}m^{0}_{s}m^{0}_{t}\Psi^{M_{d}}_{d}\rangle}{E-\frac{3}{4m}q^{2}-E_{d}}\nonumber \\
&&+\frac{\langle(-\frac{1}{2}\mathbf{p}-\frac{3}{4}\mathbf{q})(\mathbf{p}-\frac{1}{2}\mathbf{q})m_{s_{2}}m_{s_{3}}m_{s_{1}}m_{t_{2}}m_{t_{3}}m_{t_{1}}|\hat{T}|\mathbf{q}_{0}m^{0}_{s}m^{0}_{t}\Psi^{M_{d}}_{d}\rangle}{E-\frac{3}{4m}(\mathbf{p}-\frac{1}{2}\mathbf{p})^{2}-E_{d}}\nonumber \\
&&+\frac{\langle(-\frac{1}{2}\mathbf{p}+\frac{3}{4}\mathbf{q})(-\mathbf{p}-\frac{1}{2}\mathbf{q})m_{s_{3}}m_{s_{1}}m_{s_{2}}m_{t_{3}}m_{t_{1}}m_{t_{2}}|\hat{T}|\mathbf{q}_{0}m^{0}_{s}m^{0}_{t}\Psi^{M_{d}}_{d}\rangle}{E-\frac{3}{4m}(-\mathbf{p}-\frac{1}{2}\mathbf{p})^{2}-E_{d}}.\nonumber\\
\end{eqnarray}
As a simplification eq.~(\ref{Eq.13}) has been rewritten as:
\begin{eqnarray}
&&\hspace{3mm}U^{M_{d}m^{0}_{s},\gamma}_{0}(\mathbf{p},\mathbf{q};\mathbf{q}_{0})\nonumber\\&&=\frac{\hat{T}^{m^{0}_{s_{1}}m^{0}_{t_{1}},M_{d}}_{m_{s_{1}}m_{s_{2}}m_{s_{3}}m_{t_{1}}m_{t_{2}}m_{t_{3}}}(\mathbf{p},\mathbf{q};\mathbf{q}_{0})}{E-\frac{3}{4m}q^{2}-E_{d}}+\frac{\hat{T}^{m^{0}_{s_{1}}m^{0}_{t_{1}},M_{d}}_{m_{s_{2}}m_{s_{3}}m_{s_{1}}m_{t_{2}}m_{t_{3}}m_{t_{1}}}(-\frac{1}{2}\mathbf{p}-\frac{3}{4}\mathbf{q},\mathbf{p}-\frac{1}{2}\mathbf{q};\mathbf{q}_{0})}{E-\frac{3}{4m}(\mathbf{p}-\frac{1}{2}\mathbf{p})^{2}-E_{d}}\nonumber \\
&&\hspace{3mm}+\frac{\hat{T}^{m^{0}_{s_{1}}m^{0}_{t_{1}},M_{d}}_{m_{s_{3}}m_{s_{1}}m_{s_{2}}m_{t_{3}}m_{t_{1}}m_{t_{2}}}(-\frac{1}{2}\mathbf{p}+\frac{3}{4}\mathbf{q},-\mathbf{p}-\frac{1}{2}\mathbf{q};\mathbf{q}_{0})}{E-\frac{3}{4m}(-\mathbf{p}-\frac{1}{2}\mathbf{p})^{2}-E_{d}}.\nonumber\\
\end{eqnarray}

\section{Summary and outlook}\label{sec:summary}
We extended the recently developed formalism for a new treatment of
the 3N bound state in three dimensions for the Nd scattering
\cite{Bayegan}. We propose a new representation of 3D Faddeev
equation for the 3N scattering including the spin and isospin
degrees of freedom in the momentum space. This formalism stays
closely to the bosonic structure where the spin and isospin degrees
of freedom are ignored. This is an important step forward since our
3D formalism avoids the very involved angular momentum algebra
occurring for the permutations and transformations and it is more
efficient and less cumbersome for considering the 3N forces. This
formalism enables us to realistically handle more complexity in 3N
scattering calculations. This work provides the necessary formalism
for the calculation of 3N scattering observables which is under
preparation \cite{MHarzchi}.

\section*{Acknowledgments}

This work was supported by the research council of the University of
Tehran.

\appendix

\section{Connection of the deuteron wave function to its helicity representation} \label{app:t matrix1} In
In our formulation, we need the matrix elements of the deuteron wave
function
$\Psi^{M_{d}}_{m_{s_{1}}m_{s_{2}}m_{t_{1}}m_{t_{2}}}(\mathbf{p}) $.
We have connected these matrix elements to those in the
momentum-helicity basis. The momentum-helicity basis state which is
parity eigenstate and antisymmetrized is given
by~\cite{Fachruddin-PRC62}:
\begin{eqnarray}
|\textbf{p};\textbf{\^{p}}S\lambda;t\rangle^{\pi a}
&=&\frac{1}{\sqrt{2}}(1-P_{12})|\textbf{p};\textbf{\^{p}}S\lambda\rangle_{\pi}\,|t\rangle\nonumber
\\&=&\frac{1}{\sqrt{2}}(1-\eta_{\pi}(-)^{S+t})|\textbf{p};\textbf{\^{p}}S\lambda\rangle_{\pi}\,|t\rangle,
\end{eqnarray}
Here $S$ is the total spin and $\lambda$ is the spin projection
along relative momentum of two nucleons.
$|t\rangle\equiv|tm_{t}\rangle$ is the total isospin state of the
two nucleons, where $t$ is the total isospin and $m_{t}$ is the
isospin projection along its quantization axis, which tells also the
total electric charge of system. For simplicity $m_{t}$ is
suppressed since electric charge is conserved. $P_{12}$ is the
permutation operator which exchanges the two nucleons labels in all
spaces \emph{i.e.} momentum, spin and isospin spaces.
$|\textbf{p};\textbf{\^{p}}S\lambda\rangle_{\pi}$ is parity
eigenstate  which is given by:
\begin{eqnarray}
|\textbf{p};\textbf{\^{p}}S\lambda\rangle_{\pi}=\frac{1}{\sqrt{2}}(1+\eta_{\pi}P_{\pi})|\textbf{p};
\textbf{\^{p}}S\lambda\rangle,
\end{eqnarray}
where $P_{\pi}$ is parity operator, $\eta_{\pi}=\pm1$ are the parity
eigenvalues and $|\textbf{p};\textbf{\^{p}}S\lambda\rangle$ is the
momentum-helicity basis state. The normalization of the
momentum-helicity basis state is worked out as:
\begin{eqnarray}
&&\hspace{3mm}\,^{\pi'a}\langle\textbf{p}';\textbf{\^{p}}'S'\lambda';t'|\textbf{p};\textbf{\^{p}}S\lambda;\,t\rangle^{\pi
a}
\nonumber\\&&=(1-\eta_{\pi}(-)^{S+t})\,\delta_{t't}\delta_{\eta_{\pi'}\eta_{\pi}}\delta_{S'S}\{\delta(\textbf{p}'-\textbf{p})\delta_{\lambda'\lambda}+\eta_{\pi}(-)^{S}\delta(\textbf{p}'+
\textbf{p})\delta_{\lambda'-\lambda}\},
\end{eqnarray}
and the completeness relation of this state is defined by:
\begin{eqnarray}
\sum_{S\lambda\pi t}\int
d\textbf{p}\,|\textbf{p};\textbf{\^{p}}S\lambda;t\rangle^{\pi
a}\,\frac{1}{4}\ ^{\pi
a}\langle\textbf{p};\textbf{\^{p}}S\lambda;t|=1.
\end{eqnarray}
Inserting the completeness relation in the momentum helicity basis
yields:
\begin{eqnarray}\label{Eq.15}
&&\hspace{3mm}\Psi_{m_{s_{1}}m_{s_{2}}m_{t_{1}}m_{t_{2}}}^{M_{d}}(\mathbf{p})\equiv\langle
\mathbf{p}m_{s_{1}}m_{s_{2}}m_{t_{1}}m_{t_{2}}|\Psi^{M_{d}}_{d}\rangle\nonumber\\&&=\frac{1}{4}\sum_{\lambda=-1}^{1}\int
d\mathbf{p}'\langle\mathbf{p}m_{s_{1}}m_{s_{2}}m_{t_{1}}m_{t_{2}}|\textbf{p}';\textbf{\^{p}}'1\lambda;0\rangle^{1
a}\,\ ^{1
a}\langle\textbf{p}';\textbf{\^{p}}'1\lambda;0|\Psi^{M_{d}}_{d}\rangle.
\end{eqnarray}
where we have used the deuteron properties,~\emph{i.e.} $S=1$, $t=0$
and even parity. $\Phi_{\lambda}^{M_{d}}(\mathbf{p}')\equiv\
^{1a}\langle\textbf{p}';\textbf{\^{p}}'1\lambda;0|\Psi^{M_{d}}_{d}\rangle$
is the deuteron wave function component in the momentum-helicity
basis. The overlap of the momentum-helicity basis state with the
state $|\mathbf{p}m_{s_{1}}m_{s_{2}}m_{t_{1}}m_{t_{2}}\rangle$ is
given by \cite{Fachruddin-PRC62}:
\begin{eqnarray}\label{Eq.14}
&&\hspace{3mm}\langle\mathbf{p}m_{s_{1}}m_{s_{2}}m_{t_{1}}m_{t_{2}}|\textbf{p}';\textbf{\^{p}}'S\lambda;t\rangle^{\pi
a}\,\nonumber\\&&=\frac{1}{2}(1-\eta_{\pi}(-)^{S+t})\,C(\frac{1}{2}\frac{1}{2};m_{t_{1}}m_{t_{2}})\,C(\frac{1}{2}\frac{1}{2};m_{s_{1}}m_{s_{2}}\lambda_{0})\nonumber\\&&\hspace{3mm}\times
\,e^{-i\lambda_{0}\varphi}d^{S}_{\lambda_{0}\lambda}(\theta)\,[{\,\delta(\mathbf{p}-\mathbf{p}')+\eta_{\pi}\delta(\mathbf{p}+\mathbf{p}')}],
\end{eqnarray}
where $C$ is the Clebsh-Gordan coefficient and $d^{S}_{\lambda_{0}
\lambda }(\theta )$ are rotation matrices~\cite{rose}. Substituting
eq.~(\ref{Eq.14}) into eq.~(\ref{Eq.15}) yields:
\begin{eqnarray}\label{Eq.16}
&&\Psi_{m_{s_{1}}m_{s_{2}}m_{t_{1}}m_{t_{2}}}^{M_{d}}(\mathbf{p})=\frac{1}{4}\,\,C(\frac{1}{2}\frac{1}{2};m_{t_{1}}m_{t_{2}})\,C(\frac{1}{2}\frac{1}{2};m_{s_{1}}m_{s_{2}}\lambda_{0})\sum_{\lambda=-1}^{1}\Biggl\{e^{-i\lambda_{0}\varphi}d_{\lambda_{0}\lambda}^{1}(\theta)\Phi_{\lambda}^{M_{d}}(\mathbf{p})\nonumber\\&&\hspace{16mm}+e^{-i\lambda_{0}(\pi+\varphi)}d_{\lambda_{0}\lambda}^{1}(\pi-\theta)\Phi_{\lambda}^{M_{d}}(-\mathbf{p})\Biggl\}.
\end{eqnarray}
By considering these relations:
\begin{eqnarray}
d^{S}_{\lambda_{0}\lambda}(\pi-\theta)=(-)^{S+\lambda_{0}}d^{S}_{\lambda_{0}-\lambda}(\theta),\quad\quad\Phi_{\lambda}^{M_{d}}(\mathbf{p})=-\Phi_{-\lambda}^{M_{d}}(-\mathbf{p}),
\end{eqnarray}
Equation (\ref{Eq.16}) can be written as:
\begin{eqnarray}
&&\hspace{4mm}\Psi_{m_{s_{1}}m_{s_{2}}m_{t_{1}}m_{t_{2}}}^{M_{d}}(\mathbf{p})=\frac{1}{2}\,\,C(\frac{1}{2}\frac{1}{2};m_{t_{1}}m_{t_{2}})\,C(\frac{1}{2}\frac{1}{2};m_{s_{1}}m_{s_{2}}\lambda_{0})\,e^{-i\lambda_{0}\varphi}\sum_{\lambda=-1}^{1}d_{\lambda_{0}\lambda}^{1}(\theta)\Phi_{\lambda}^{M_{d}}(\mathbf{p}).\nonumber\\
\end{eqnarray}
The azimuthal dependency of $\Phi_{\lambda}^{M_{d}}(\mathbf{p})$ is:
\begin{eqnarray}
\Phi_{\lambda}^{M_{d}}(\mathbf{p})\equiv
e^{iM_{d}\varphi}\Phi_{\lambda}^{M_{d}}(p,\theta)\end{eqnarray}
Finally the connection between the
$\Psi^{M_{d}}_{m_{s_{1}}m_{s_{2}}m_{t_{1}}m_{t_{2}}}(\mathbf{p}) $
and those in the momentum-helicity basis state is given by:
\begin{eqnarray}
\Psi_{m_{s_{1}}m_{s_{2}}m_{t_{1}}m_{t_{2}}}^{M_{d}}(\mathbf{p})=\frac{1}{2}\,\,C(\frac{1}{2}\frac{1}{2};m_{t_{1}}m_{t_{2}})\,C(\frac{1}{2}\frac{1}{2};m_{s_{1}}m_{s_{2}}\lambda_{0})\,e^{-i(\lambda_{0}-M_{d})\varphi}\sum_{\lambda=-1}^{1}d_{\lambda_{0}\lambda}^{1}(\theta)\Phi_{\lambda}^{M_{d}}(p,\theta).\nonumber\\
\end{eqnarray}
It should be mentioned that the deuteron wave function component
$\Phi_{\lambda}^{M_{d}}(p,\theta)$ obeys a set of coupled equations
which are solved numerically in ref. \cite{Fachruddinnh}.

\section{Appendix B. Anti-symmetrized NN $\textbf{t}$-matrix and connection to its helicity
representation} \label{app:t matrix2}

 The connection of Anti-symmetrized two-body $t$-matrix to those in
the momentum-helicity basis is given in
ref.~\cite{Fachruddin-PRC68}, here we prepare this connection
according to the notation to be used in our work and then we have
derived an expansion in momentum-helicity basis for two-body
$t$-matrix for $z\longrightarrow E_{d}$. Based on momentum-helicity
basis states the NN $t$-matrix element is defined as:
\begin{eqnarray}
t_{\lambda \lambda '}^{\pi St}({{\bf p}, \, {\bf p}' };z) \equiv \,
^{\pi a} \langle {\bf p};\hat{{\bf p}}S\lambda ;t | t(z)|{\bf
p}';\hat{{\bf p}}'S\lambda' ;t \rangle ^{\pi a}.
\end{eqnarray}
As shown in ref.~\cite{Fachruddin-PRC68}, the selection of ${\bf
p}'$ parallel to the $z$-axis allows, together with the properties
of the potential, that the angular dependencies of the NN $t$-matrix
elements can be simplified as:
\begin{eqnarray}
t_{\lambda \lambda '}^{\pi St}({{\bf p}, \, {\bf p}' };z) &=& e^{-i
\lambda \Omega_{pp'} } \,\, t_{\lambda \lambda '}^{\pi St}({p{\,\bf
\hat{n}}_{pp'}, \, p'{\bf \hat{z}} };z) \nonumber
\\* &=&
e^{-i \lambda \Omega_{pp'} } \,\, e^{i \lambda' \phi_{pp'} } \,
t_{\lambda \lambda '}^{\pi St}(p, p', \cos \theta_{pp'} ;z)
\nonumber
\\* &\equiv&
 e^{i(\lambda '\phi_{pp'} -\lambda \Omega_{pp'} )} \,\, t_{\lambda
\lambda '}^{\pi St}(p,p',\cos \theta_{pp'} ;z),
\end{eqnarray}
The direction ${\bf \hat{n}}_{pp'}$ can be determined by the
spherical and the polar angles $\theta _{pp'}$ and $\varphi_{pp'}$,
where
\begin{eqnarray}
\cos \theta_{pp'} &=&  \cos \theta_{p} \cos \theta_{p'} +\sin
\theta_{p} \sin \theta_{p'} \cos (\phi_{p} -\phi_{p'}), \nonumber
\\*
\sin \theta_{pp'} e^{i \varphi_{pp'}} &=&  -\cos \theta_{p} \sin
\theta_{p'} +\sin \theta_{p} \cos \theta_{p'} \cos (\phi_{p}
-\phi_{p'}) + i \sin \theta_{p} \sin (\phi_{p} -\phi_{p'}),
\end{eqnarray}
and the exponential factor $ e^{i(\lambda' \phi_{pp'} -\lambda
\Omega )} $ is calculated as:
\begin{eqnarray}
e^{i\lambda \Omega_{pp'} }  &=& \frac{\sum ^{S}_{N=-S}
D^{S}_{N\lambda }( \phi_{p} \, \theta_{p} \, 0 ) D^{\ast
S}_{N\lambda '} (\phi_{p'} \, \theta_{p'} \, 0 )}{D^{S}_{\lambda
'\lambda }( \phi_{pp'} \, \theta_{pp'} \, 0)}, \nonumber
\\*
e^{i(\lambda '\phi_{pp'} -\lambda \Omega_{pp'} )}  &=& \frac{\sum
^{S}_{N=-S}e^{iN(\phi_{p} -\phi_{p'} )}d^{S}_{N\lambda }(\theta_{p}
)d^{S}_{N\lambda '} (\theta_{p'} )}{d^{S}_{\lambda '\lambda
}(\theta_{pp'})}. \nonumber\\
\end{eqnarray}
In the above expressions, $D^{S}_{N\lambda }( \phi_{p} \, \theta_{p}
\, 0 )$ are the Wigner D-functions. Finally the connection between
the $t$-matrix elements $ _{a}\langle{\bf p} \, m_{s_{1}} m_{s_{2}}
\, m_{t_{1}} m_{t_{2}} |\hat{t}(z)| {\bf p}' \, m'_{s_{1}}
m'_{s_{2}} \, m'_{t_{1}} m'_{t_{2}} \rangle _{a}$ and those in the
momentum-helicity basis, namely $t_{\lambda \lambda '}^{\pi St}({\bf
p},{\bf p}';z)$, is given as:
\begin{eqnarray}
&&\hspace{3mm} _{a}\langle{\bf p} \, m_{S} m_{S} \, m_{t_{1}}
m_{t_{2}} |\hat{t}(z)| {\bf p}' \, m'_{s_{1}} m'_{s_{2}} \,
m'_{t_{1}} m'_{t_{2}} \rangle _{a}  \nonumber \\* && =\frac{1}{4} \,
\delta _{m_{t_{1}} +m_{t_{2}},m'_{t_{1}}+m'_{t_{2}}}(z-E_{d})
\nonumber \\* &&\hspace{3mm}\times e^{-i(\lambda _{0}\phi _p
-\lambda _{0}'\phi _{p' })} \, \sum _{S\pi t}( 1-\eta _{\pi
}(-)^{S+t}) \nonumber
\\* && \hspace{3mm} \times
C(\frac{1}{2}\frac{1}{2}t; m_{t_{1}} m_{t_{2}} )
C(\frac{1}{2}\frac{1}{2}t; m'_{t_{1}} m'_{t_{2}} ) \nonumber
\\* && \hspace{3mm} \times C(\frac{1}{2}\frac{1}{2}S; m_{s_{1}} m_{s_{2}} )
C(\frac{1}{2}\frac{1}{2}S; m'_{s_{1}} m'_{s_{2}} ) \nonumber
\\* && \hspace{3mm} \times
\sum _{\lambda \lambda '}d^{S}_{\lambda _{0}\lambda }(\theta
_p)d^{S}_{\lambda _{0}'\lambda '}(\theta _{p' })t_{\lambda \lambda
'}^{\pi St}({{\bf p}, \, {\bf p}' };z). \label{eq.t_a-t_helicity}
\end{eqnarray}
It should be mentioned that the fully off-shell two-body $t$-matrix
$t_{\lambda \lambda '}^{\pi St}(p,p',\cos \theta_{pp'} ;z)$ obeys a
set of coupled Lippman-Schwinger equations which for \(S = 0\) it is
a single equation and for \(S = 1\) it is a set of two coupled
equations which are solved numerically in ref. \cite{Fachruddinn}.

In this stage we have derived an expression in the momentum-helicity
basis for NN $t$-matrix for $z\longrightarrow E_{d}$. It is clear
that:
\begin{eqnarray}
\lim_{z\longrightarrow
E_{d}}(z-E_{d})t=V|\Psi_{d}^{M_{d}}\rangle\langle\Psi_{d}^{M_{d}}|V.
\end{eqnarray}
Projecting this equation on momentum-helicity basis states yields:
\begin{eqnarray}
&&\hspace{3mm}\lim_{z\longrightarrow
E_{d}}(z-E_{d})t^{110}_{\lambda\lambda'}(\mathbf{p},\mathbf{p}')\nonumber\\&&=\,^{1a}\langle\mathbf{p};\mathbf{\hat{p}}1\lambda;0|V|\Psi_{d}^{M_{d}}\rangle^{1a}\,\,^{1a}\langle\Psi_{d}^{M_{d}}|V|\mathbf{p}';\mathbf{\hat{p}}'1\lambda';0\rangle^{1a}\nonumber\\&&=
\Biggl\{\frac{1}{4}\sum_{\lambda''}\int d
\mathbf{p}''\,^{1a}\langle\mathbf{p};\mathbf{\hat{p}}1\lambda;0|V|\mathbf{p}'';\hat{\mathbf{p}}''1\lambda'';0\rangle^{1a}\,\,^{1a}\langle\mathbf{p}'';\hat{\mathbf{p}}''1\lambda'';0|\Psi_{d}^{M_{d}}\rangle\Biggl\}\nonumber\\&&\hspace{3mm}\times\Biggl\{\frac{1}{4}\sum_{\lambda''}\int
d
\mathbf{p}''\langle\Psi_{d}^{M_{d}}|\mathbf{p}'';\hat{\mathbf{p}}''1\lambda'';0\rangle^{1a}\,\,^{1a}\langle\mathbf{p}'';\hat{\mathbf{p}}''1\lambda'';0|V|\mathbf{p}';\mathbf{\hat{p}}'1\lambda';0\rangle^{1a}\Biggl\}\nonumber\\&&=
\Biggl\{\frac{1}{4}\sum_{\lambda''}\int d
\mathbf{p}''V^{110}_{\lambda\lambda''}(\mathbf{p},\mathbf{p}'')\,\Phi^{M_{d}}_{\lambda''}(\mathbf{p}'')\Biggl\}\Biggl\{\frac{1}{4}\sum_{\lambda''}\int
d
\mathbf{p}''V^{110}_{\lambda''\lambda'}(\mathbf{p}'',\mathbf{p}')\,\Phi^{\ast
M_{d}}_{\lambda''}(\mathbf{p}'')\Biggl\}.
\end{eqnarray}
By considering vector $\mathbf{q}'$ along $z$ axis we have obtained:
\begin{eqnarray}\label{Eq.17}
&&e^{i\lambda'\varphi}\lim_{z\longrightarrow
E_{d}}(z-E_{d})t^{110}_{\lambda\lambda'}(p,p',x)\nonumber\\&&=
\Biggl\{\frac{1}{4}\sum_{\lambda''}\int d
\mathbf{p}''e^{iM_{d}\varphi''}V^{110}_{\lambda\lambda''}(\mathbf{p},\mathbf{p}'')\,\Phi^{M_{d}}_{\lambda''}(p'',x'')\Biggl\}\nonumber\\&&\hspace{3mm}\times\Biggl\{\frac{1}{4}\sum_{\lambda''}\int
d
\mathbf{p}''e^{-iM_{d}\varphi''}e^{i\lambda'\varphi''}V^{110}_{\lambda''\lambda'}(p'',p',x'')\,\Phi^{
M_{d}}_{\lambda''}(p'',x'')\Biggl\},
\end{eqnarray}
where we have used the azimuthal behavior of the potential and the
deuteron wave function as:
\begin{eqnarray}
V^{\pi
St}_{\lambda''\lambda'}(\mathbf{p}'',p'\,\hat{\mathbf{z}})&=&e^{i\lambda'\varphi''}V^{\pi
St}_{\lambda''\lambda'}(p'',p',x''),\nonumber\\\Phi^{M_{d}}_{\lambda''}(\mathbf{p}'')&=&e^{iM_{d}\varphi''}\Phi^{M_{d}}_{\lambda''}(p'',x'')
\end{eqnarray}
Equation (\ref{Eq.17}) can be rewritten as:
\begin{eqnarray}
&&\hspace{3mm}\lim_{z\longrightarrow
E_{d}}(z-E_{d})t^{110}_{\lambda\lambda'}(p,p',x)\nonumber\\&&=
\Biggl\{\frac{1}{4}\sum_{\lambda''}\int d
\mathbf{p}''e^{-iM_{d}(\varphi-\varphi'')}V^{110}_{\lambda\lambda''}(\mathbf{p},\mathbf{p}'')\,\Phi^{M_{d}}_{\lambda''}(p'',x'')\Biggl\}\nonumber\\&&\hspace{3mm}\times\Biggl\{\frac{1}{4}\sum_{\lambda''}\int
d
\mathbf{p}''e^{-iM_{d}(\varphi''-\varphi)}e^{i\lambda'(\varphi''-\varphi)}\,V^{110}_{\lambda''\lambda'}(p'',p,x'')\,\Phi^{
M_{d}}_{\lambda''}(p'',x'')\Biggl\}\nonumber\\&&=
\Biggl\{\frac{1}{4}\sum_{\lambda''}\int_{0}^{\infty}dp''p''^{2}
\int_{-1}^{1}
dx''\upsilon^{110,M_{d}}_{\lambda\lambda''}(p,x,p'',x'')\,\Phi^{M_{d}}_{\lambda''}(p'',x'')\Biggl\}\nonumber\\&&\hspace{3mm}\times\Biggl\{\frac{1}{4}\sum_{\lambda''}\int_{0}^{\infty}dp''p''^{2}
\int_{-1}^{1}
dx''\,V^{110}_{\lambda''\lambda'}(p'',p',x'')\,\Phi^{M_{d}}_{\lambda''}(p'',x'')\int_{0}^{2\pi}d
\varphi''e^{i(\lambda'-M_{d})(\varphi''-\varphi)}\Biggl\},\nonumber\\
\end{eqnarray}
where we have introduced:
\begin{eqnarray}
\upsilon^{110,M_{d}}_{\lambda\lambda''}(p,x,p'',x'')=\int_{0}^{2\pi}d\varphi''e^{-iM_{d}(\varphi-\varphi'')}V^{110}_{\lambda\lambda''}(\mathbf{p},\mathbf{p}''),
\end{eqnarray}
Finally we have obtained:
\begin{eqnarray}
&&\hspace{3mm}\lim_{z\longrightarrow
E_{d}}(z-E_{d})t^{110}_{\lambda\lambda'}(p,p',x)\nonumber\\&&=
\Biggl\{\frac{1}{4}\sum_{\lambda''}\int_{0}^{\infty}dp''p''^{2}
\int_{-1}^{1}
dx''\upsilon^{110,M_{d}}_{\lambda\lambda''}(p,x,p'',x'')\,\Phi^{M_{d}}_{\lambda''}(p'',x'')\Biggl\}\nonumber\\&&\hspace{3mm}\times\Biggl\{\frac{\pi}{2}\,\delta_{\lambda'M_{d}}\sum_{\lambda''}\int_{0}^{\infty}dp''p''^{2}
\int_{-1}^{1}
dx''\,V^{110}_{\lambda''\lambda'}(p'',p',x'')\,\Phi^{M_{d}}_{\lambda''}(p'',x'')\,\Biggl\},
\end{eqnarray}

\end{document}